\documentclass[oldversion]{aa}
\usepackage{graphicx}
\usepackage{natbib}
\usepackage{graphicx}
\usepackage{revsymb}	
\usepackage{amsmath}	
\usepackage[usenames]{color}	
\usepackage{epstopdf}	
\newcommand{\inv} {\frac {1}}

\newcommand{\derivp} [2] {\frac {\partial #1 } {\partial #2} }
\newcommand{\deriv} [2] {\frac {\textrm{d} #1 } {\textrm{d} #2} }

\newcommand{\eq}[1] {Eq.\,(\ref{#1})}
\newcommand{\eqn} [1] {
\begin{equation} #1
\end{equation}}

\usepackage{txfonts}
\usepackage[applemac]{inputenc}

\begin{document}

\title{Solar-like oscillations in massive main-sequence stars}
\subtitle{I. Asteroseismic signatures of the driving and damping regions}

 \author{K. Belkacem\inst{1}, M.A. Dupret\inst{1}, and A. Noels\inst{1}} 

\institute{Institut d’Astrophysique et de G\'eophysique, Universit\'e de Li\`ege, All\'ee du 6 Ao\^ut 17-B 4000 Li\`ege, Belgium }

   \offprints{K. Belkacem}
   \mail{Kevin.Belkacem@obspm.fr}
   \date{\today}

  \authorrunning{Belkacem et al.}
  \titlerunning{Solar-like oscillations in massive main-sequence stars}

  \abstract{Motivated by the recent detection of stochastically excited modes in the massive star V1449 Aql  (Belkacem et al., 2009b), already known to be a $\beta$ Cephei, we theoretically investigate the driving by turbulent convection. By using a full non-adiabatic computation of the damping rates, together with a computation of the energy injection rates, we provide an estimate of the amplitudes of modes excited by both the convective region induced by the iron opacity bump and the convective core. Despite uncertainties in the dynamical properties of such convective regions, we demonstrate that both are able to efficiently excite $p$ modes above the CoRoT observational threshold and the solar amplitudes. In addition, we emphasise the potential asteroseismic diagnostics provided by each convective region, which we hope will help to identify the one responsible for solar-like oscillations, and to give constraints on this convective zone. A forthcoming work will be dedicated to an extended investigation of the likelihood of solar-like oscillations  across the Hertzsprung-Russell diagram.}


   \maketitle
   
\section{Introduction}
\label{intro}

$\beta$ Cephei-type stars are known to pulsate with large amplitude oscillations \citep{Gautschy95a,Gautschy95b}. The $\kappa$-mechanism is at the origin of those large amplitude oscillations and is related to the iron opacity bump at $\log T \approx 5.3$ \citep[e.g.][]{Pamyatnykh99}. Those modes are linearly unstable. 
In contrast, low-amplitude modes have been observed in the Sun for years and are thought to be intrinsically stable but stochastically excited by turbulent convection. 
Their excitation is attributed to turbulent convection and takes place in the uppermost
part of the Sun, which is a place of vigourous and turbulent motions. Since the pioneering work of
\citet{Lighthill52}, we know that a turbulent medium generates
incoherent acoustic pressure fluctuations (also called acoustic ``noise''). 
In the last decade, solar-like oscillations have been detected in
numerous F-G type main sequence and red giant stars, in different evolutionary stages and with different
metallicities \citep[see the recent review by][]{Bedding07}. 

As in the Sun, massive stars have convective regions, namely the convective core, and two regions associated with the helium and iron opacity bumps. 
The inner convective region (\emph{i.e.} the convective core) has an important impact on the internal structure and on the subsequent evolution of the star through, for instance, the extent of the  region where chemical elements are mixed \citep[e.g.][]{Kippenhahn90}. 
In contrast, the external convective zones associated with opacity bumps are generally thought to be unimportant since they transport only a very small fraction of the energy flux, which is dominated by radiation. It is only since the revision of the opacities  \citep{Iglesias92}, which led to an enhancement of the iron opacity bump and could thus account for the instability in $\beta$ Cephei stars, that a convective zone related to the iron opacity bump has been thought to be present in massive stars. 
Much attention is now drawn to this convective region associated with the iron opacity bump\footnote{In the following, this convective region will be referred to as \emph{iron convective region}.}, because it is of  interest for the understanding of surface effects such as microturbulence or wind clumping \citep[e.g.][]{Cantiello09}. 
However, both the convective core and the iron convection zone are poorly understood. In particular, the typical length-scale and time-scale of turbulent motions associated with these regions are only inferred through mixing-length arguments, as well as dynamical properties such as the convective velocities. 

Solar-like oscillations (linearly damped and stochastically excited modes) are closely related to those dynamical properties, so their detection in such massive stars is a promising way to infer them. 
Until recently, stochastically excited modes have only been predicted and detected in solar-like and red giant stars \citep[see][for details]{Bedding07,Michel08,DeRidder09}. The data gathered by the CoRoT mission allowed us to report the first detection of solar like oscillations in a massive star, V1449 Aql \citep{Belkacem09b}. Those oscillations exhibit mode amplitudes higher than solar, hence the issue of their excitation is of interest.  Investigation of unstable modes already gives constraints on the internal structure of massive stars \citep[e.g.][]{Aerts03,Thoul03,Dupret04,Pamyatnykh04} as well as on the underlying excitation mechanisms \citep[e.g.][]{Moskalik92,Dziembowski93a,Dziembowski93b,Dziembowski08}. The discovery of solar-like oscillations is thus an opportunity to go a step further in the understanding of those stars which still challenge theory \citep{Dziembowski07}.

Our objective in this work is to assess the excitation of solar-like oscillations in massive main-sequence stars, such as $\beta$ Cephei type stars, by turbulent convection. We aim at determining which convective region is able to excite modes above the observational threshold, as well as emphasising the possible asteroseismic diagnostic that would permit us to identify the excitation region and to obtain constraints on this region.  
As the amplitude of stochastically excited modes is a balance between driving and damping, both must be modelled. In this work we will use the formalism proposed by 
\cite{Samadi00I} and extended by \cite{Belkacem06a,Belkacem06b,Belkacem08}.  
Damping rates are computed using the MAD code \citep{MAD02,MAD05}, which is based on a full non-radial non-adiabatic treatment. 
This first paper of the series focuses on a benchmark model. An extended investigation of the mode amplitude through the Hertzsprung-Russell diagram will be the object of the second paper of this series. 

The paper is organised as follows: Sect.~2 introduces the general
formalism that is used to compute mode amplitudes.  
In Sect.~3, we focus on mode damping rates and we emphasise their frequency dependence. 
Excitation by the convective region associated with the iron convective region is also studied 
and a possible diagnostic is inferred. Section~4 presents the computation of mode amplitude excited by the convective core. Conclusions are formulated in Sect.~5.

\section{Modelling mode amplitudes}

\subsection{Equilibrium stellar structure; a $10\, M_\odot$ model as a benchmark}
\label{model_1D}

The objective is to explore the potential of convective regions to drive oscillations. As a benchmark we investigate the driving of radial acoustic modes by turbulent convection zones in a $10\, M_\odot$ main-sequence star, which is typical of $\beta$ Cephei stars \citep{Pamyatnykh99}. 
A more systematic exploration of the HR diagram is dedicated to the second paper of this series. 

\subsubsection{Physical input}
\label{phys_input}

The stellar structure models used in this work are 
obtained with the stellar evolution code CL\'ES \citep{CLES}. 
In the interior model, we used the OPAL
opacities \citep{Opal96} extended to low temperatures with the
opacities of \cite{Alexander94} and the CEFF equation of state
\citep{CD1992}. The metallicity is assumed to be $Z=0.02$ with the 
metal mixture as derived by \cite{GN93}. 
Convection is included according to 
a B\"{o}hm-Vitense mixing-length (MLT) formalism and no 
overshoot is allowed.  
The effect of the mixing-length parameter is   
discussed in Sects.~\ref{ampl_iron} and \ref{ampl_core}. 
 Note, however, that the default value of the mixing-length is $1.8$ (the Solar value) and 
in the convective region associated to the iron opacity bump, the mixing-length is of the same order that 
the extent of the convective region, $1.5 \%$ of the star radius.

\begin{figure}
\begin{center}
\includegraphics[height=6cm,width=9cm]{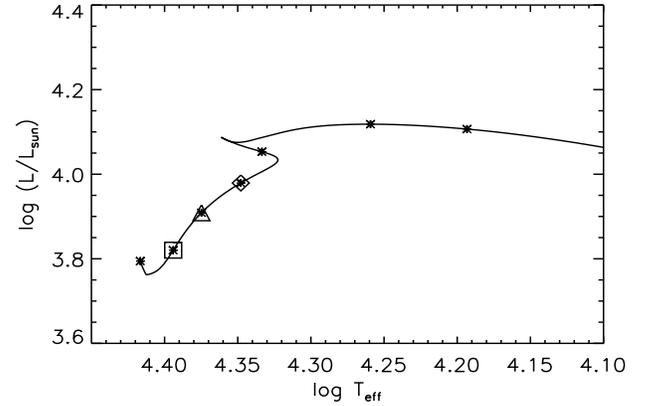}
\caption{Hertzsprung-Russell diagram for a ten solar mass model computed with the CL\'ES code (see Sect.~\ref{model_1D}). The diamond corresponds to the benchmark model as described in Sect.~\ref{phys_input}, the triangle symbols corresponds to another model used for comparison in Sect.~\ref{diag_iron}.  Star symbols correspond to the different evolutionary stages displayed in Fig.~\ref{fig4} (bottom). 
Eventually, the square and triangle symbols are used to identified the models presented in Fig.~\ref{fig4} (top).}
\label{fig0}
\end{center}
\end{figure}

The location of the benchmark model on the Hertzsprung-Russell diagram is plotted in 
Fig.~\ref{fig0}. The model has an effective temperature of $\log T_{\rm eff}= 4.35$, and a surface gravity of $\log g = 3.80$. 

\subsubsection{Convective zones}
\label{conv_zones}

\begin{figure}
\begin{center}
\includegraphics[height=6cm,width=9cm]{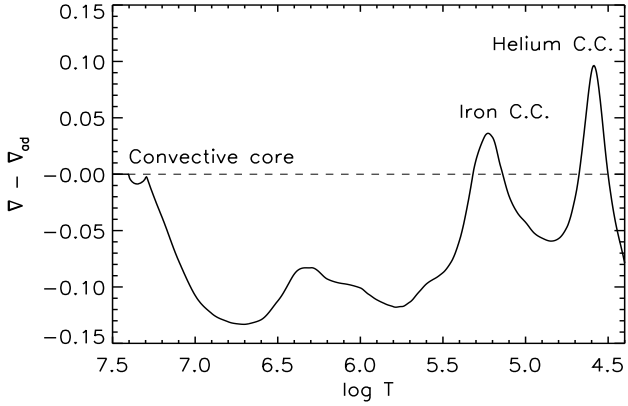}
\caption{The difference between the temperature gradient and the adiabatic one ($\nabla - \nabla_{ad}$) versus the temperature. In the convective core, the difference $\nabla - \nabla_{ad}$ vanishes since convection is efficient and transport almost all the energy flux. In contrast, in the outer convective regions,  we see significant gradients which means that convective transport is not efficient there.}
\label{fig00}
\end{center}
\end{figure}

In such massive stars, one can distinguish three convective regions, namely; 
\begin{itemize}
\item the convective zone associated with the \emph{helium II} opacity bump. This region is located at $T\approx 40\,000$ K (Fig.~\ref{fig00}), and is very weak in the sense that it is inefficient in transporting energy with a negligible ratio of the convective heat flux to the radiative one. In addition, such a region is located near the stellar surface, where the density is low, and velocities are in the $10$ to $20$ m.s$^{-1}$ range (low heat capacity is the cause of effective smoothing of the temperature contrast and reduction of the buoyant acceleration). Hence, the kinetic energy that could feed modes is small.   
\item the convective zone associated with the \emph{iron} opacity bump, located at $T\approx 200\,000 K$ (Fig.~\ref{fig00}). 
 It has been shown \citep{Cantiello09} that the occurrence of convection in the iron opacity peak is strongly dependent on the luminosity and the metallicity of the star. At a given luminosity, a smaller metallicity implies a lower opacity, which means that a larger fraction of the total flux can be transported by radiation. On the other hand, increasing the luminosity at a given $Z$ leads to a more important contribution of convection in the energy transport. For a given metallicity, there is a luminosity threshold (of the order of $10^{3.2} \,L_{\odot}$ for $Z=0.02$) above which a convective zone is associated with the iron peak. 

As for the helium convective region, the transport of energy by convection is very low. However, this region is located deeper than the helium one, where density is higher, 
with convective velocities up to $5.5$ km.s$^{-1}$. The kinetic energy flux is still negligible compared to the total flux but the ratio of the kinetic energy flux of the iron convective zone to the helium one is $\approx 8 \times 10^7$.
\item the \emph{convective core}. Convection is fully adiabatic (fig.~\ref{fig00}), and a significant part of  the total flux is transported by convection. The kinetic energy is four order of magnitudes higher than for the iron convective region, due to high densities, with velocities around $300$  m.s$^{-1}$.
\end{itemize}
We conclude that the convective zone associated with the helium opacity bump is unlikely to be able to efficiently excite solar-like oscillations. Indeed, it has been verified numerically that the resulting theoretical mode amplitudes are found well below the observational detection threshold of CoRoT (around one ppm). In contrast, the convective zones associated with the iron opacity bump as well as with the convective core are more promising. The available kinetic energy flux is important. However, to be efficient, mode excitation must fulfil other criteria, which are discussed in Sects.~\ref{iron} and \ref{core}.  

\subsection{Computation of mode amplitudes}
\label{comp_amplitude}
 
We compute the mean-squared surface velocity for each radial mode as described in \cite{Belkacem06b,Belkacem09}, \emph{i.e.}
\begin{align}
\label{Pobs}
v_s^2 (h) = \frac{ \mathcal{P}}{2 \, \eta \, \mathcal{M} } 
\end{align}
where $\mathcal{M}=I/|\xi_r(h)|^2$ denotes the mode mass, $I$ the mode inertia, $\eta$ the damping rate, $\mathcal{P}$ the energy injection rate, and $h$ is the height in the stellar atmosphere. 
In this section, we consider the level of the photosphere $h=R$ with $R$ being the radius 
at the photosphere. 

The mode amplitude in terms of intensity is then deduced at the photosphere, as proposed by 
\cite{Dziembowski77a} and \cite{Pesnell90}, \emph{i.e.}
\begin{equation}
\frac{\delta L}{L} = 4\, \frac{\delta T_{\rm eff}}{T_{\rm eff}} + 2\, \frac{\xi_r}{r}
\end{equation}
where $\delta L$ is the bolometric mode intensity fluctuation, and $\delta T_{\rm eff}$ the mode effective temperature fluctuation. The relation between the variation of effective temperature and the mode 
displacement is obtained from MAD \citep[see][for details]{MAD02,Dupret02b}. 
Eventually, as the mode amplitude is a balance between the driving ($\mathcal{P}$) and the damping ($\eta$), both are to be determined and computed.

\subsubsection{Computation of the damping rates}
\label{comp_damping}

As shown by \eq{Pobs}, the computation of mode amplitudes requires the knowledge of both energy injection ($\mathcal{P}$) and damping ($\eta$) rates. The latter has been computed using the full non-adiabatic 
pulsation code MAD \citep{MAD02}. This code includes a time-dependent convection 
(TDC) treatment described in \cite{MAD05}, which allows us to take into account the role played 
by the variations of the convective flux, the turbulent pressure, and the dissipation rate of 
turbulent kinetic energy. 

The damping of acoustic modes is found to be dominated by the perturbation of the radiative flux. 
Note that, although crucial for the excitation, the upper convective regions do not affect the damping rates for stars we are considering. Indeed, convection is very inefficient for the transport of energy (compared to the radiative transport) and the feedback of the convective background on the pulsation remains small. We have numerically verified that the interaction between convection and pulsation does not affect   damping rates as well as the effect of turbulent pressure and the dissipation rate of turbulent kinetic energy. 

Hence, the dominant contribution to the damping rate can be written
\begin{equation}
\label{dampings_radiatif}
\eta = \frac{1}{2 \, \omega_0 I} \int_{0}^{M} \mathcal{I}m \left[
 \left(\frac{\delta \rho}{\rho_0}^* T_0 \delta S\right) \left(\Gamma_3 - 1\right) \right] \textrm{d}m
\end{equation}
where $\delta S$ is the perturbation of entropy, $\delta \rho$ the perturbation of density, $T_0$ the mean temperature, $\omega_0$ the mode frequency, $\rho_0$ the mean density, $m$ the local mass, $M$ the total mass, and 
the star stands for the complex conjugate. 

Keeping only the radial contribution of the radiative flux in the energy equation 
because of its dominant contribution, and neglecting the production of nuclear energy ($\delta \epsilon=0$), 
one gets
\begin{align}
\label{energy}
i \sigma T_0 \delta S = - \deriv{\delta L_r}{m} 
\end{align}
where $\sigma = \omega_0 + i \eta$, and $L_r$ is the radiative flux. 
In addition, in the diffusion approximation, one can write 
\begin{equation}
\label{deltaL}
\frac{\delta L}{L} =  \left( - \frac{\delta \kappa}{\kappa} +
\frac{1}{\left(\textrm{d}T / \textrm{d}r\right)}\frac{\partial \delta T}{\partial r} 
+2\,  \frac{\xi_r}{r} + 3\, \frac{\delta T}{T} 
 - \frac{\delta \rho}{\rho} 
- \frac{\partial \xi_r}{\partial r}\right) \, .
\end{equation}
Finally, in the quasi-adiabatic approximation (which is valid in the deep layers) one has
\begin{equation}
\label{delta_kappa}
\frac{\delta \kappa}{\kappa} \approx \kappa_{p s} \frac{\delta p}{p} \qquad \mbox{with}\qquad 
\kappa_{p s} = \left(\kappa_\rho + \left(\Gamma_3 - 1\right) \kappa_T\right) / \Gamma_1
\end{equation}
and
\begin{align}
\kappa_{p s} = \left( \derivp{\ln \kappa}{\ln p} \right)_s \, ,
\kappa_\rho =  \left( \derivp{\ln \kappa}{\ln \rho} \right)_T \, ,
\kappa_T = \left( \derivp{\ln \kappa}{\ln T} \right)_\rho
\end{align}

\subsubsection{ Energy injection rates ($\mathcal{P}$)}
\label{excitation}

The formalism we use to compute energy injection rates of radial modes 
was developed by \cite{Samadi00I} and extended by \cite{Belkacem06a,Belkacem06b}. 
It takes two sources into account, the Reynolds stress tensor and 
the advection of the turbulent fluctuations of entropy by
the turbulent motions (the ''entropy source term''). 
The entropy contribution is found to be around ten percent 
of the Reynolds one.  

The energy injection rate, $\mathcal{P}$, then mainly arises from the Reynolds stresses  and can 
be written  as \citep[see Eq.~(21) of][]{Belkacem08} 
\begin{align}
\label{puissance}
\mathcal{P} &=\frac{16}{15}  \frac{\pi^{3}}{2  I} \int_{0}^{M}  \textrm{d}m   \, \rho_0  \, \left|\deriv{\xi_r}{r}\right|^2 \; \mathcal{S}_{\rm R} 
\end{align}
with
\begin{align}
\label{puissance2}
\mathcal{S}_{\rm R} &=  \int_{0}^{+\infty}  \frac{\textrm{d}k}{k^2 }~E^2(k) \int_{-\infty}^{+\infty} \textrm{d}\omega  
~\chi_k( \omega + \omega_0) ~\chi_k( \omega )
\end{align}
where 
$\mathcal{S}_{\rm R}$ is the source function,  $E(k)$ the spatial kinetic energy 
spectrum, $\chi_k$ the eddy-time correlation function, and $k$ the wave-number. 

The rate ($\mathcal{P}$) at which energy is injected into a mode is then 
computed according to \eq{puissance}. Eigenfrequencies and eigenfunctions  are computed
using the non-adiabatic pulsation code MAD. 
In addition to the eigenfunctions and the density stratification, 
\eq{puissance} involves both the convective velocity and the turbulent kinetic energy  spectrum.
The typical convective length-scales are poorly known for main sequence massive stars. 
Hence, the classical mixing-length theory is used to find the injection 
length-scale (\emph{i.e.} the scale at which the 
turbulent kinetic energy spectrum is maximum) and 
a parameter $\beta$ is introduced \citep[see][for details]{Samadi00I}
such that the associated wave-number is $k_0 = 2\pi / \beta \,\Lambda$,
where $\Lambda$ is the mixing-length. 
One has also to specify the way that turbulent eddies are temporally-correlated by defining 
an eddy-time correlation function. A Lorentzian function 
had successfully been used; in the solar case \citep{Samadi02I,Belkacem06b}, 
as well as for $\alpha$ Cen A, and HD49933 \citep{Samadi08,Samadi09}, it indeed reproduces 
the observational data. 
Consequently, if not explicitly mentionned, such modelling will be used  in the following.

\section{Driving by the iron convective region}
\subsection{Damping rates}
\subsubsection{Efficiency of the damping}

\begin{figure}
\begin{center}
\includegraphics[height=6cm,width=9cm]{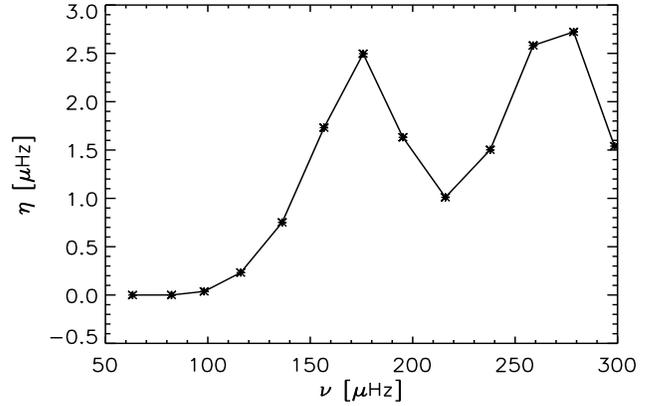}
\caption{Damping rates ($\eta$) of radial modes as a function of frequency, computed as described 
in Sect.~\ref{comp_damping} using the stellar model introduced in Sect.~\ref{model_1D}. The fundamental radial mode is found unstable ($\eta < 0$) while the overtones are found stable ($\eta > 0$).}
\label{fig1}
\end{center}
\end{figure}

In Fig.~\ref{fig1}, the fundamental mode is found linearly unstable with respect to the $\kappa$-mechanism associated with the iron opacity bump \citep{Pamyatnykh99}. Higher radial-order modes are found linearly stable. The damping rates range between one and three $\mu$Hz for $\nu > 150\, \mu$Hz, while they nearly vanish around the unstable mode. They are found to be dominated by the first two terms of \eq{deltaL}. 

Concerning the first contribution ($\delta \kappa / \kappa$), 
as described for instance by \cite{Pamyatnykh99}, the region where $\kappa_\rho + \left(\Gamma_3 - 1\right) \kappa_T$ increases outwards tends to drive the mode while the region where $\kappa_\rho + \left(\Gamma_3 - 1\right) \kappa_T$ decreases outward damps the mode. This contribution then mainly determines the damping rates of low-order modes while the second term of \eq{deltaL} begins to be important for higher order modes.

\begin{figure}
\begin{center}
\includegraphics[height=6cm,width=9cm]{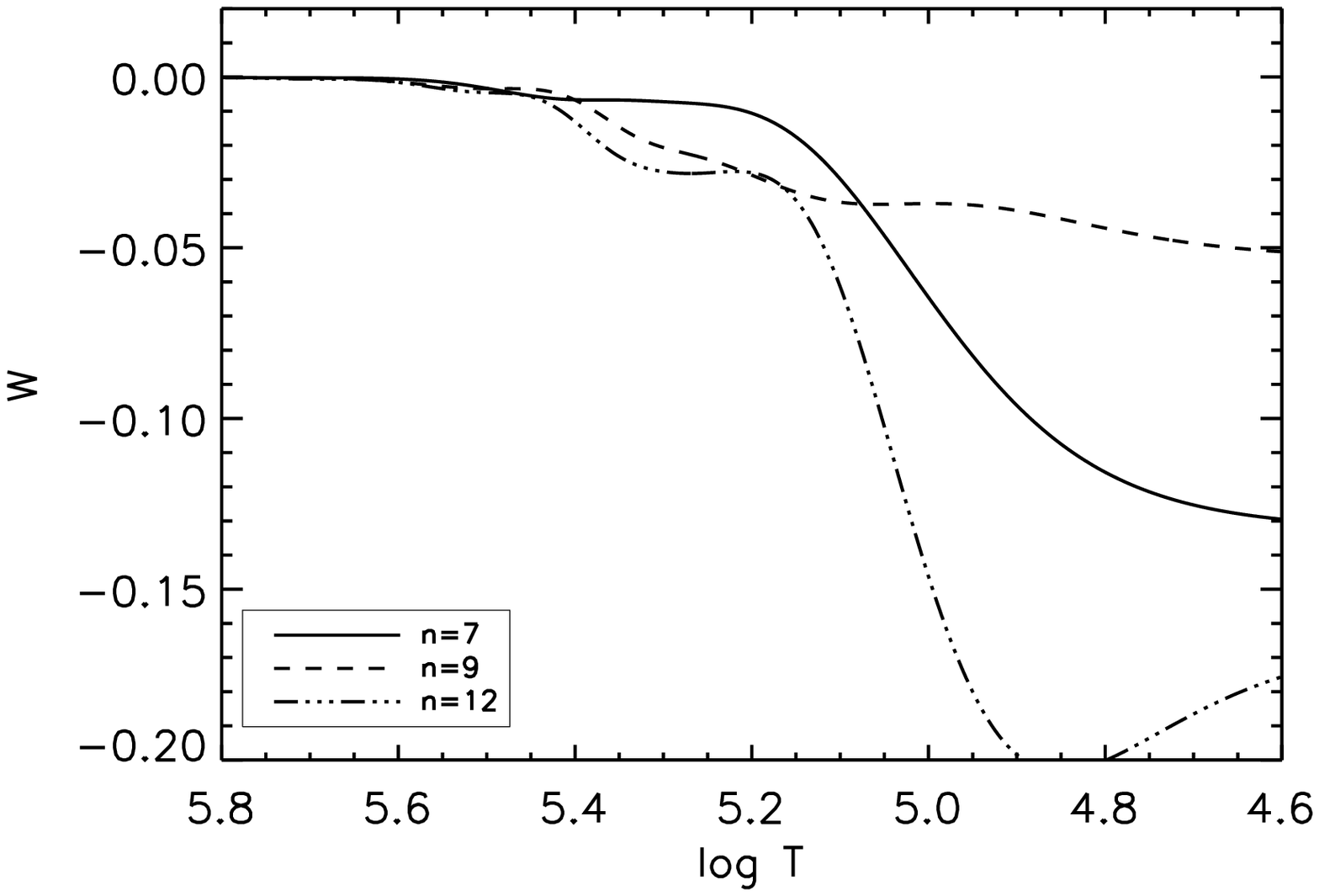}
\includegraphics[height=6cm,width=9cm]{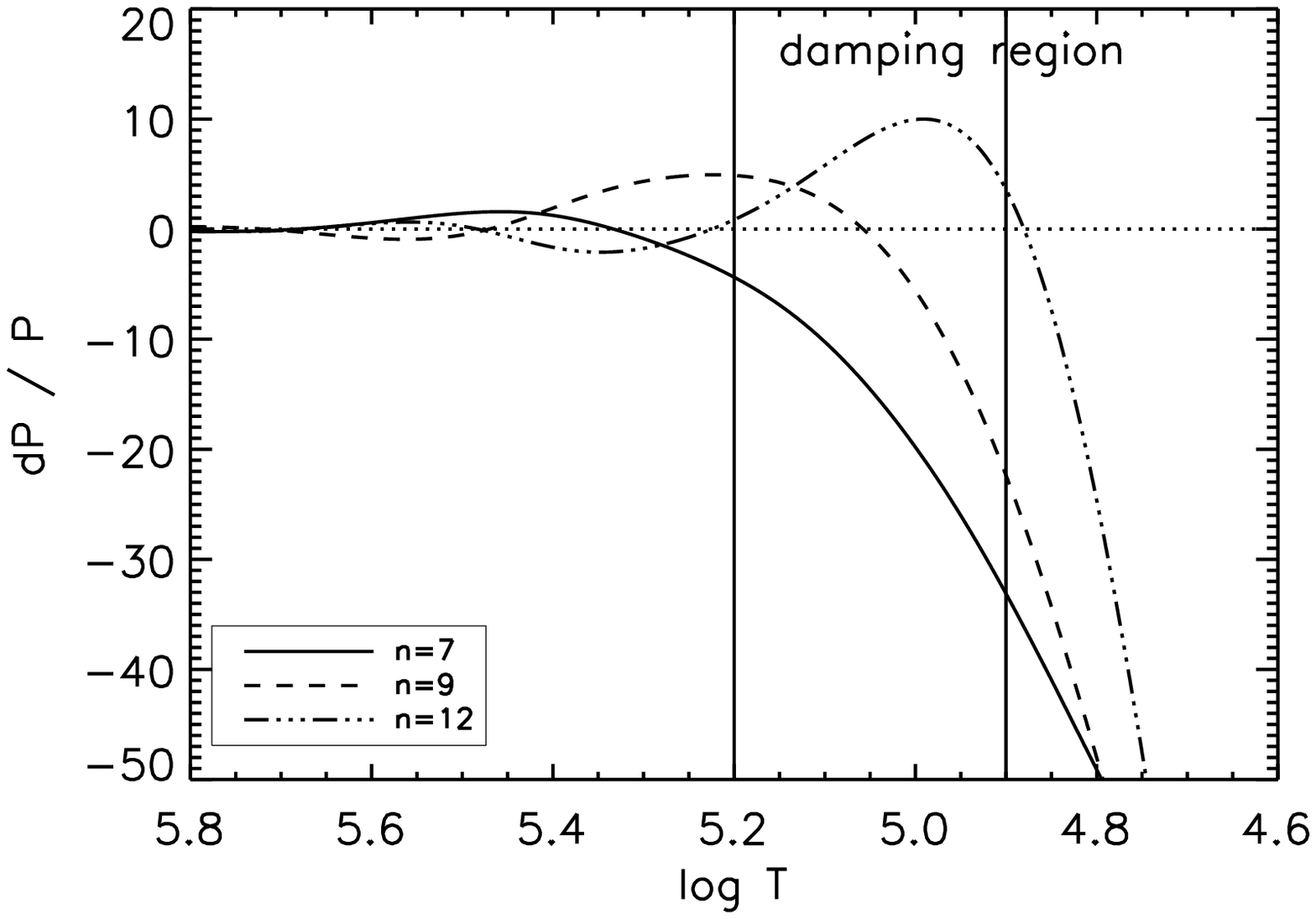} 
\caption{{\bf Top:} Cumulative work integral integrals ($W$) versus the logarithm of the temperature for the $n=7$, $n=9$ and $n=12$, $\ell=0$ modes. An increases outward of $W$ corresponds to a driving while  a decrease inward, to a damping. The surface values give the final values of the work integral. By convention, the mode is found stable if the surface value of $W$ is negative and unstable if positive. 
{\bf Bottom:} Real part of the perturbation of pressure ($\delta P/P$) related to the $n=7$, $n=9$ and $n=12$, $\ell=0$ modes versus the logarithm of the temperature. The horizontal dots lines permits to visualise the nodes of the eigen modes, and the vertical lines delimit the damping region.}
\label{fig2}
\end{center}
\end{figure}

\subsubsection{Periodic variations}

For high frequency modes, the damping in the range $\log T \in [5.2;4.9]$  dominates over the driving region ($\log T \in [5.5;5.3]$). This can be seen in Fig.~\ref{fig2}, in which the work integral is plotted, for three  modes which correspond to the first and the second maxima and the minimum in between of the damping rate (see Fig.~\ref{fig1}). 

The oscillation of the damping rates for high-frequency modes ($\nu > 175 \, \mu$Hz) is related to the location of the node of $\delta P / P$ with respect to the damping region. 
By considering the pressure fluctuations ($\delta P / P$) of three modes, namely $n=7$, $n=9$ and $n=12$, Figure \ref{fig2} shows that the damping is minimum when there is a node in the damping region while maximum when the nodes are located at the edges of those layers. The second maximum  arises when two nodes of $\delta P / P$ straddle the damping region. 

This is easily explained by the set of equations \eq{dampings_radiatif}-\eq{delta_kappa}, since a node in pressure fluctuations lowers the opacity fluctuations that appear to be the main damping term in \eq{deltaL}.  Note also that the second derivative of $\delta T / T$ is negative when $\delta T / T$ is positive, which strengthens the radiative losses. 

It then creates the periodic variations observed in Fig.~\ref{fig1} for $\nu > 175 \, \mu$Hz. Hence, the detection of such an oscillation of mode line-width (damping) would be a signature of the iron opacity bump and also an opportunity to get constraints on modes pressure perturbation in this region. 

\subsection{ Energy injection rates}
\label{iron}

The poor knowledge of the dynamical properties of convective regions associated with the iron opacity bump leads us to use a simple description based on the MLT as explained in Sect.~\ref{excitation}. In terms of  energy injection rates, two major features are to be determined: the convective velocity and the injection length-scale. 
As for the superficial convective layers of the Sun, the iron convection region is inefficient since it transports a negligible part of the energy flux 
through convection. The Mach number ($\mathcal{M} \approx 0.1$) is similar to that found in the upper convective layers of the Sun in which $\mathcal{M} \approx 0.3$. 
To what extent one can use the values of the parameters obtained for the Sun is however difficult to assess  and only numerical simulations will provide a firm answer.  
By default, for the iron convective region, we will use the values of the convective velocity provided by MLT  using the assumed solar value ($\alpha=1.8$). In addition, the injection length scale is deduced from the numerical simulations of the upper part of the solar convection zone as derived by \cite{Samadi02I}, \emph{i.e.} using $\beta = 5$ (see Sect.~\ref{excitation}).

\subsubsection{Efficiency of the excitation}

A common feature of every modelling of stochastic excitation by turbulent convection is that the energy injection rate is locally proportional to the kinetic energy flux \citep[e.g.,][]{GK94,Samadi00I}. The contribution due to the Reynolds stresses can be simplified as \citep{Samadi09}
\begin{equation}
\label{approx_P}
\mathcal{P} \propto \, \left(\frac{\omega_0}{c_s}\right)^2 \, \mathcal{F}_{\rm kin} \, \Lambda^4 
\end{equation}
where $\Lambda$ is a characteristic length, $c_s$ is the sound speed, and $\mathcal{F}_{\rm kin}$ is the specific kinetic energy flux defined as
\begin{equation}
\mathcal{F}_{\rm kin} = \rho \, u_{\rm rms}^3
\end{equation}
with $\rho$ the density, and $u_{\rm rms}$ the root mean square velocity given by the MLT. 

The helium convective region is very inefficient in exciting modes since  the kinetic energy flux is small (see Sect.~\ref{conv_zones}). The more favourable convective regions, in this star, are the convective core and the iron opacity bump induced one. 
Excitation of radial acoustic modes by the iron convective region is found to be efficient for two reasons.  
First, the iron opacity bump is located deeper in the star compared to the helium bump. 
 Both, density and turbulent velocities are higher. Thus, the energy available for $p$ modes is higher.  Second, the efficiency of the excitation depends on the involved time-scales, 
\emph{i.e.} the convective time-scale and the modal period. 
The latter is several hours and, using mixing-length arguments (\emph{i.e.} $\tau \approx \Lambda / u_{\rm mlt}$), we find the convective time-scale to be also of several hours. Hence, excitation is nearly resonant. 

\begin{figure}[!]
\begin{center}
\includegraphics[height=6cm,width=9cm]{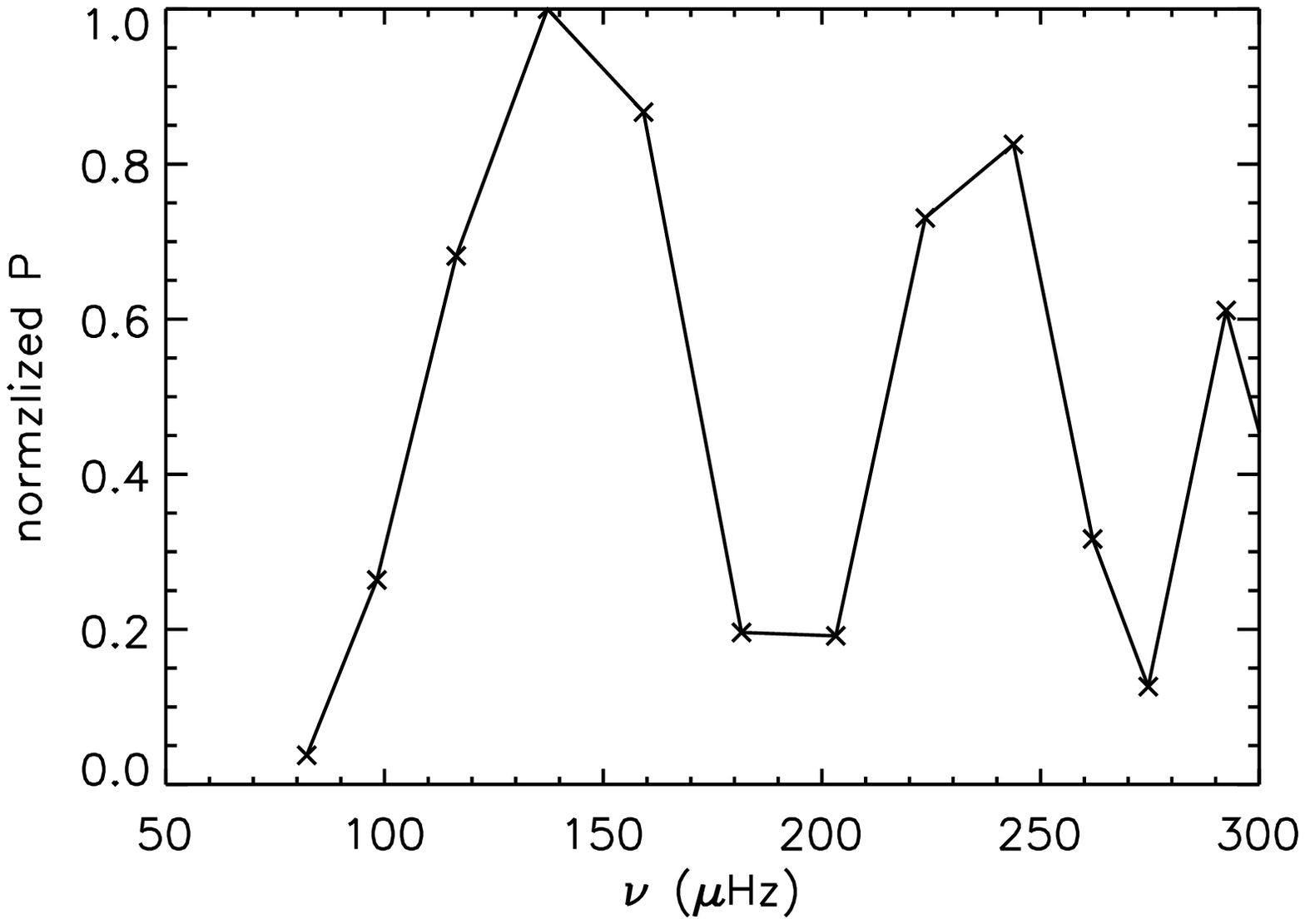}\\
\includegraphics[height=6cm,width=9cm]{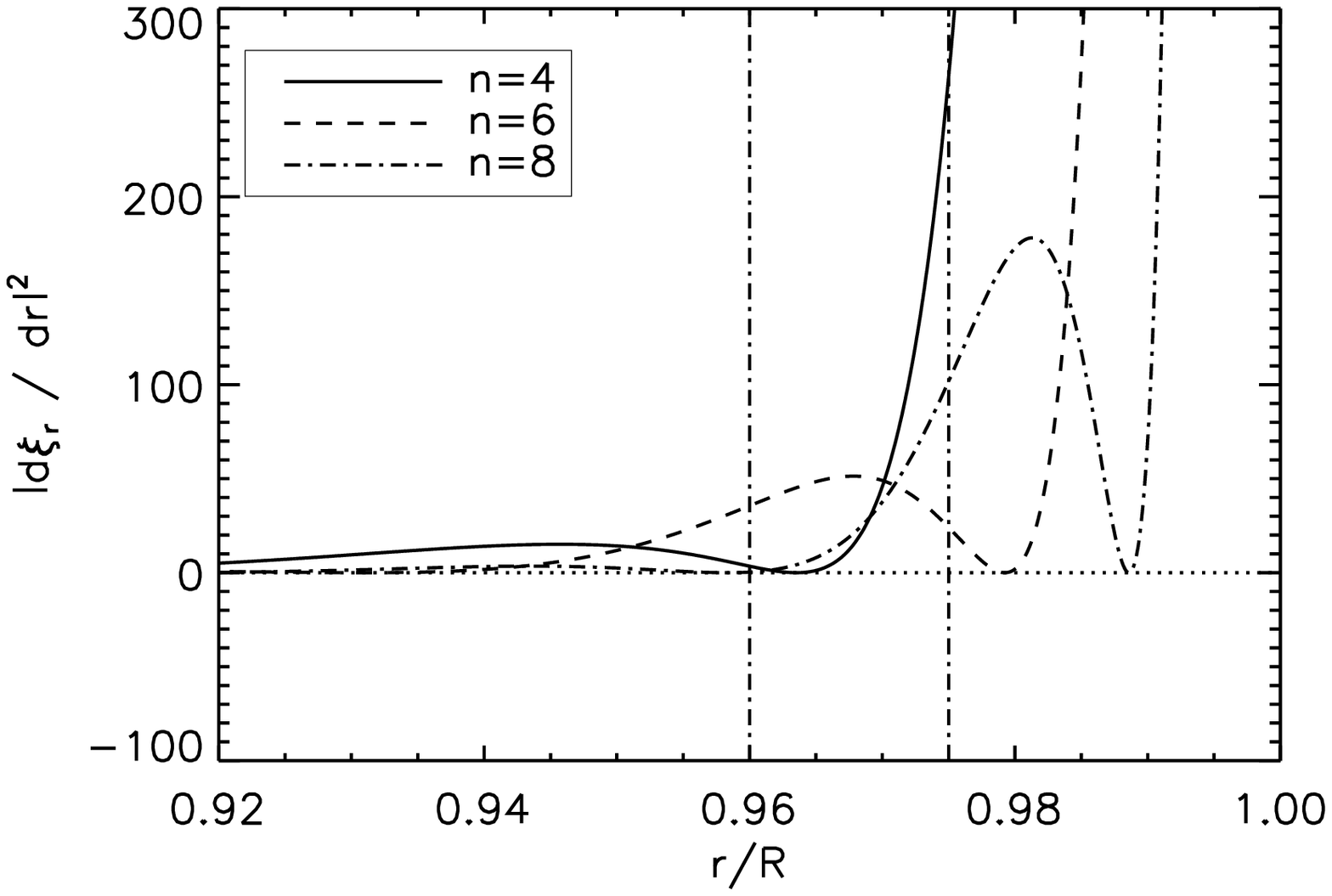}
\caption{{\bf Top:} Rates ($\mathcal{P}$) at which energy is supplied to the modes, computed as detailed 
in Sect.~\ref{excitation}, for a $10 \, M_\odot$ main-sequence model (see Sect.~\ref{model_1D}). Note that only the contribution of the iron convective region is included in $P$.   
{\bf Bottom:}  Square of the radial derivative of the radial component of the eigenfunction for radial orders $n=4, n=6$ and $n=8$ that correspond in the top panel of the first maximum, minimum and the second maximum of the energy injection rates. The vertical lines delimit the location of the convective region induced by the iron opacity bump.}
\label{fig3}
\end{center}
\end{figure}

\subsubsection{Periodic variation and seismic diagnostic}
\label{diag_iron}

The power supplied to the mode by the iron convection zone exhibits periodic variations with frequency (Fig.~\ref{fig3}, top). Such a behaviour is explained by the location of the iron convection zone compared to the radial component of the displacement. As shown by \eq{puissance}, the energy injection rate depends on the radial derivative of the radial component of the displacement in the driving regions. Hence, the location of the displacement  nodes relatively to the iron convective zone determines its derivative, subsequently the efficiency of the excitation. 
Such an oscillation is therefore a signature of the excitation by the iron convective region. Observational evidence would lead to associating the driving with this region since, as will be shown in Sect.~\ref{core}, no similar results are obtained if we assume that the convective core drives the modes. 

The location of the iron convective zone can be inferred from the energy injection rates by identifying the radial order of the first maximum, which corresponds to the mode whose last node is located at the bottom of the convective region (see Fig.~\ref{fig3}, bottom). A second way to proceed is to use the frequencies of two the maxima. 
The successive maxima of the energy injection rates satisfy the relation
\begin{equation}
\int_{r_{b}}^{R} k_r {\rm d}r^\prime + \phi = m \, \pi \; ,
\end{equation}
where $m$ is an integer, $k_r$ the radial wave-number, $\phi$ the phase, $R$ is the star radius, $r_b$ is the location of the node which we identify as an upper limit of the depth of the iron convective region. Hence, by considering the frequencies of two maxima of the energy injection rates (denoted by the subscripts $1$ and $2$), the location of the bottom of the iron convective region can be deduced from the integral expression
\begin{equation}
\label{eq_kr}
\int_{r_{b}}^{R} k_{r 2} {\rm d}r^\prime -  \int_{r_{b}}^{R} k_{r 1} {\rm d}r^\prime = \pi  
\end{equation}

From \eq{eq_kr}, one immediately sees that only the frequencies of two consecutive maxima are needed to determine the depth of the bottom of the iron convective region. To illustrate this, we consider the limit of pure radial acoustic waves, such that \eq{eq_kr} becomes
\begin{equation}
\left( \omega_2 - \omega_1 \right) \int_{r_{b}}^{R} \frac{{\rm d}r^\prime}{c_s} = \pi \, .
\end{equation}
Furthermore, in the very crude simplification where the sound speed is constant, one obtains the very simple analytic relation 
\begin{align}
\label{approx_Lp}
R-r_b = \frac{\pi c_s}{\left( \omega_1 - \omega_2 \right)}
\end{align}
Using the benchmark model described in Sect.~\ref{model_1D}, one has $c_s\approx 40$ km.s$^{-1}$ and from Fig.~\ref{fig3} we have $\omega_1 \approx 150 \, \mu$Hz and $\omega_2 \approx 250 \, \mu$Hz. 
From \eq{approx_Lp} we get $(R-r_b)/R \approx 0.044$, while from Fig.~\ref{fig3} one has $(R-r_b)/R = 0.04$. This order of estimate then illustrates that the detection of solar-like oscillations excited by the iron convective region could give information on the structure of the iron convective region.

\begin{figure}[!]
\begin{center}
\includegraphics[height=6cm,width=9cm]{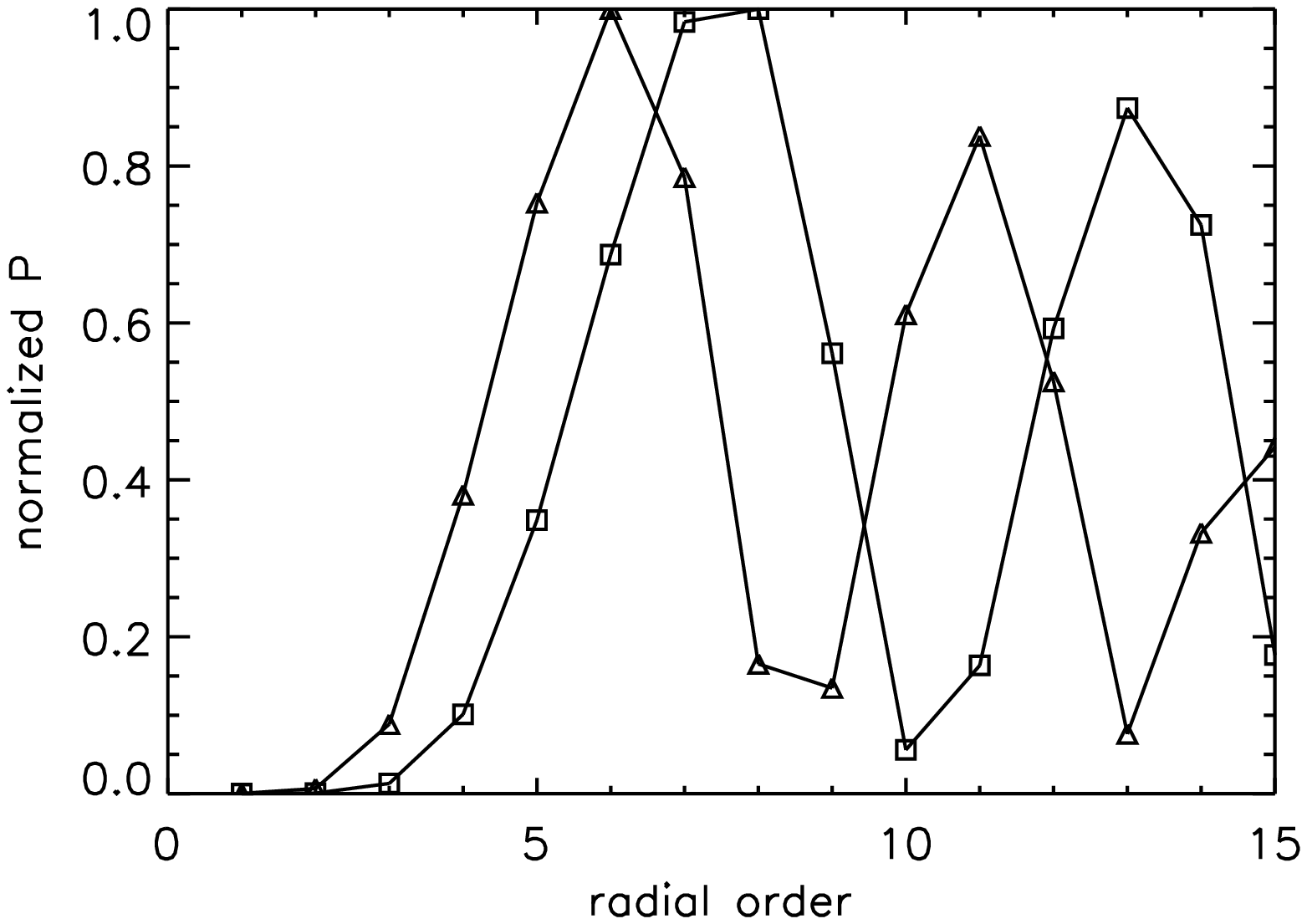}\\
\includegraphics[height=6cm,width=9cm]{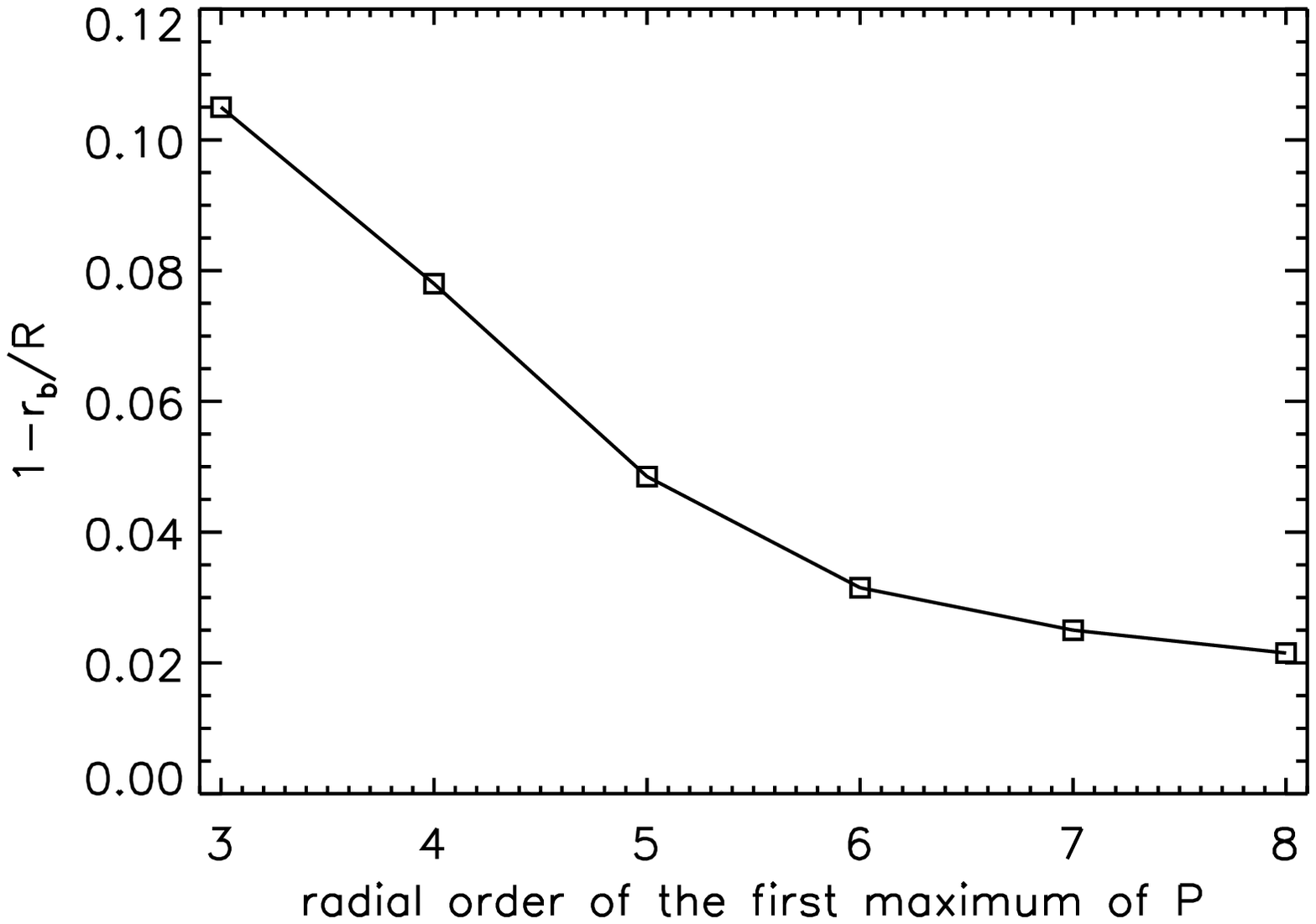}
\caption{{\bf Top:} Normalised energy injection rates for two evolutionary stages (the two first successive models in Fig.~\ref{fig0}) as a function of the mode radial order. {\bf Bottom:} Location of the bottom of the iron convection zone, with respect to the surface, versus the radial-order of the first maximum of the energy injection rates. 
Each point corresponds to the 10 $M_\odot$ modes described in Sect.~\ref{model_1D} at different evolutionary stages displayed in Fig.~\ref{fig0}.} 
\label{fig4}
\end{center}
\end{figure}

To illustrate this point, figure~\ref{fig4} (top) displays the energy injection rates as function of the radial order for two models at different evolution stages. The more evolved the star is, the deeper the iron convective zone is located since the effective temperature decreases as the star evolves. 
It turns out that the modes for which the energy injection rate is maximum shift toward lower radial-order. As a results, it is possible to follow and localise the depth of the convective region (see Fig.~\ref{fig4}, bottom). 
Note that energy injection rates are sensitive to the location of the bottom of the convective zone to less than one percent of the star radius, as shown in  Fig.~\ref{fig4}.

\subsection{Mode amplitude}
\label{ampl_iron}

\begin{figure}[!]
\begin{center}
\includegraphics[height=6cm,width=9cm]{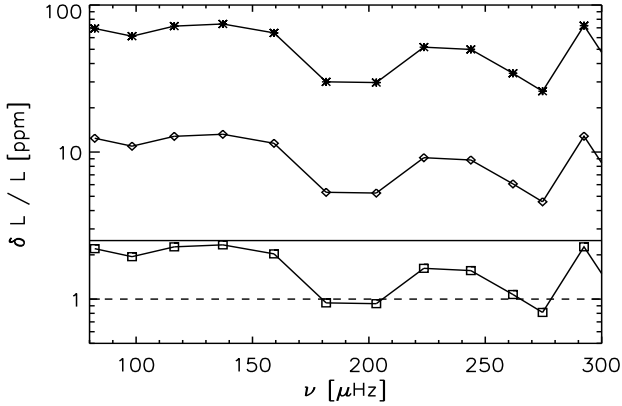}
\caption{Mode amplitude as a function of mode frequency, computed as described in Sect.~\ref{excitation} and \ref{comp_amplitude}. Diamonds correspond to the computation of the energy injection rates using $\beta=5$ and the velocities given by the MLT, while for stars the velocities have been multiplied by a factor two and divided by two for squares (note that a factor two corresponds to a variation of $30 \%$ of the mixing-length). 
The solid horizontal line corresponds to the maximum for the Sun as derived by \cite{Michel09}, and the dashed horizontal line to the CoRoT detection threshold.}
\label{fig5}
\end{center}
\end{figure}

Figure~\ref{fig5} displays mode amplitudes excited by the iron convective region. It appears that using the same values of the parameters as in the Sun, the amplitudes are found above the solar maximum and the CoRoT detection threshold. In addition, Fig.~\ref{fig5} also shows that those results are very sensitive to the convective velocities. It shows that a factor two in the velocities leads to important discrepancies for mode amplitudes, so does the mixing length, since the convective velocities depend on the mixing-lentgh to the third power for inefficient convection. 

The development of numerical simulations is then the only way to obtain more reliable values for  mode amplitudes.

\section{Driving by core convection}
\label{core}

In contrast with the uppermost convective layers of the star, the inner convective core transports a non-negligible part of the total energy flux; convection is highly efficient. Hence, one cannot infer properties of those regions by simply looking for similarities with the solar convective region. Indeed, the physics is different and the dynamical properties, which are of interest to compute mode amplitude, are very uncertain. For instance, in the adiabatic regime the mixing-length has no significant impact on the determination of the star convective flux. The typical length-scale and the convective velocities are, however, of crucial importance to determine the efficiency of the driving. 

Recent 3D numerical simulations can give some hints on those properties for massive stars. A recent study of convection in the efficient regime has been proposed by \cite{Meakin07} using compressible 3D numerical simulations. They investigated convective regions during oxygen shell burning and during hydrogen core burning for a 23 $M_\odot$ star and proposed an extended description of the convective properties and comparison with mixing-length theory as well as other 3D numerical simulations of convective outer layers. 
It turns out that both the velocities and the typical-length-scale are in rather good agreement with mixing-length-theory with a parameter ($\alpha$) of around $1.7$.

\begin{figure}[!]
\begin{center}
\includegraphics[height=6cm,width=9cm]{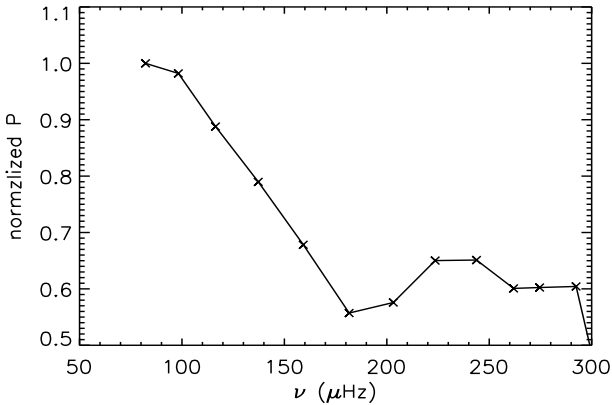}
\caption{Rates ($\mathcal{P}$) at which energy is supplied to the modes, computed as detailed 
in Sect.~\ref{excitation}, for our benchmark model, by the convective core. }
\label{fig6}
\end{center}
\end{figure}

\subsection{ Energy injection rates}

As already pointed out in Sect.~\ref{conv_zones}, the kinetic energy flux in the convective core is an order of magnitude higher than in the iron convective region. Such a large flux is due to the large densities and then makes the convective core a good candidate to excite modes. 
 However, the excitation efficiency depends also on the matching between convective time-scales and mode periods and on the shape of eigenfunctions (see Eqs.~\ref{puissance} and \ref{puissance2}).

Figure.~\ref{fig6} displays the normalised energy injection rates as a function of frequency for radial $p$ modes. 
One can distinguish between two regions, namely low $\nu \, (\in [70;200] \, \mu$Hz) and high  $\nu \, (\in [200;300] \, \mu$Hz) frequencies.
In the low frequency range, the decrease of the energy injection rates is dominated by the decrease of the ratio between the convective turn-over time scale and the modal period, and also the shape of the eigenfunctions since $\mathcal{P}$ depends on its radial derivative. In contrast, modes with higher frequencies exhibit nodes near the interface between the convective and radiative zones explaining the modulation of energy injection rates. In terms of absolute values for the energy injection rates, as for the iron convection zone, they depend mainly on the injection length-scale but are not very sensitive to the mixing-length since  convection is efficient 
(see Sect.~\ref{ampl_core}). 

\subsection{Diagnostic on the eddy-time correlation function}

\begin{figure}[!]
\begin{center}
\includegraphics[height=6cm,width=9cm]{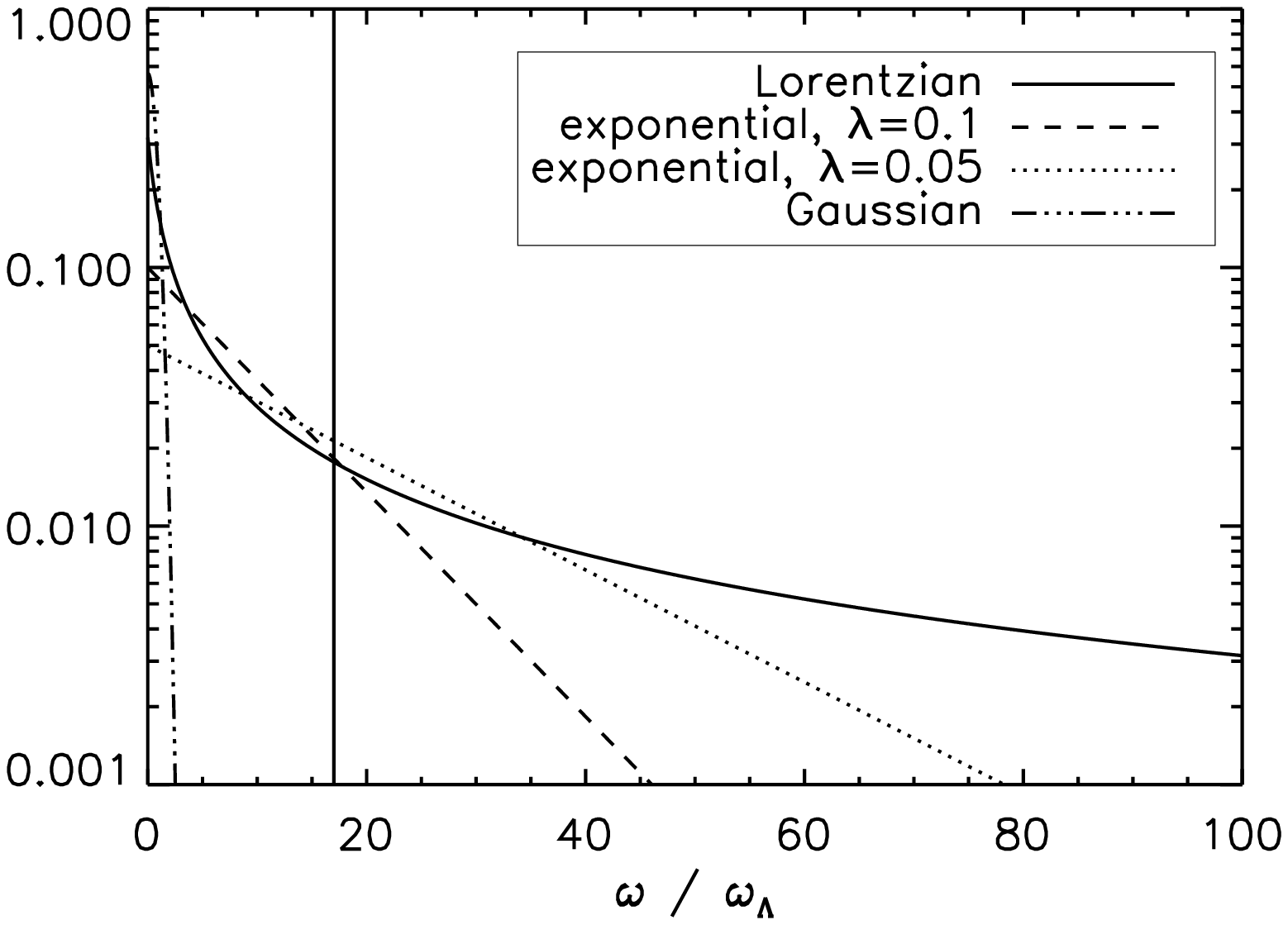}
\includegraphics[height=6cm,width=9cm]{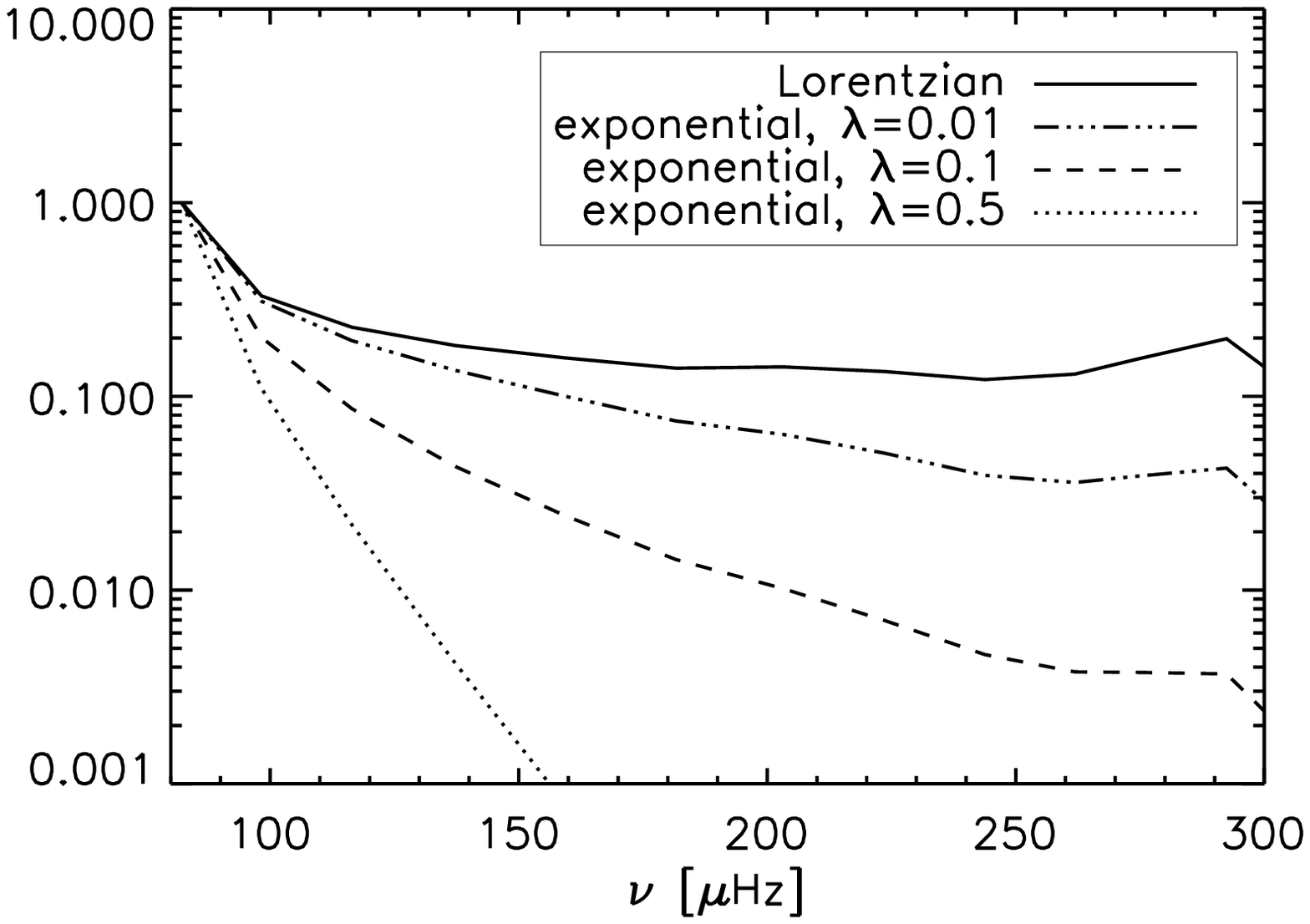}
\caption{{\bf Top:} Analytical eddy-time correlation function ($\chi_k$) as a function of the ration $\omega / \omega_k$, where $\omega_k$ is the convective frequency at the wave-number $k$. The vertical line  corresponds to the ratio $\omega / \omega_k$ for the most energetic eddies and for the mode $n=2$. Hence for higher frequencies, the ratio  $\omega / \omega_k$ will increase, and the driving will be in the off resonance regime. 
{\bf Bottom:}. Normalized mode amplitude computed using the eddy-time correlation functions as described in the upper panel. }
\label{fig7}
\end{center}
\end{figure}

Convective turn-over time-scales, evaluated in the MLT frame, are found to be of the 
order of a month, while the modal periods of acoustic standing waves are several hours. 
Hence, the excitation will be in an off resonance regime and will therefore crucially depends on the 
way eddies are time-correlated. In this particular regime, the way the eddies are temporally-correlated 
can lead to order of magnitude differences in terms of mode amplitudes, as shown by \cite{Samadi02II} for the solar $p$ modes and by \cite{Belkacem09} for solar $g$ modes, where the effect is much stronger. 

Therefore, we consider different modelling of this time-correlation function (see $\chi_k$ in \eq{puissance}). Exponential and Lorentzian modelling reads 
\begin{align}
\chi_k (\omega ) &= \inv  { \omega_k}  \mathrm{e}^{-|\omega / \omega_k|} \label{eqn:GF}\\
\chi_k (\omega ) &=\frac{1} {\pi \omega_k} \,\frac{1}{1+ \left( \omega / \omega_k \right )^2}
\label{eqn:GF1}
\end{align}
with the normalisation condition
\begin{equation}
\int_{-\infty}^{+\infty} \chi_k (\omega ) \textrm{d}\omega = 1
\end{equation}
where $\omega_k$ is the linewidth, defined as
\eqn{
\omega_k  \equiv \,  { \,k \, u_k \over \lambda}  
\; .
\label{eqn:omegak}
} 
where $\lambda$ is a parameter introduced to account for the uncertainties in defining $\omega_k$ \citep[][]{B92a,Samadi00I,Chaplin05},   
the velocity  $u_k$ of the eddy with wavenumber $k$ 
is related to the kinetic  energy spectrum $E(k)$ by   \citep{Stein67} 
\eqn{
u_k^2 =  \int_k^{2 k}  {\rm d}k \, E(k) \; .
\label{eqn:uk2}
} 
The choice of an exponential function is motivated by experimental studies of homogeneous 
and isotropic turbulence. For instance, \cite{Mordant04} have demonstrated that the time-correlation 
of lagrangian velocities follows an exponential decrease in agreement with the Kolmogorov 
(1941) phenomenology.  
However, the solar turbulent convection exhibits different physical conditions compared to experimental studies. 
The very large Reynolds numbers as well as the presence of coherent large-scale structures (plumes) make it likely that a specific description of $\chi_k$ is needed. Indeed, a Lorentzian description can be  adopted and has proven to better reproduce both the observations \citep{Belkacem06b} and the 3D numerical simulations for both $p$ and $g$ modes \citep{Samadi02II,Belkacem09}. 

Figure~\ref{fig7} (top) displays analytical $\chi_k$ versus the ratio $\omega / \omega_k$, where $\omega_k$ is the eddies characteristic frequency at the wave-number $k$. It turns out that by considering the most energetic eddies, excitation of $p$ modes will occur in the off resonance regime with a ratio $\omega / \omega_k$ higher than around 20. In that regime, it appears that the choice of $\chi_k$ is essential as well as the width $\omega_k$. Note also that a Gaussian modelling leads to a vanishing driving since for such ratios $\omega / \omega_k$, the correlation is almost zero. 
Figure~\ref{fig7} (bottom) also shows the influence of the frequency behaviour of the mode amplitudes with frequency. The shape of the eddy-time correlation function has a great impact on the shape of mode amplitudes. 
A Lorentzian function, which slowly decreases with frequency, results in a relatively small decrease of mode amplitude with increasing frequency. In contrast,  an exponential $\chi_k$ leads to a more steep  slope for mode amplitude that makes the detection less likely. 

In conclusion, we have shown that the efficiency of the driving by core convection depends crucially on the way the eddies are temporally-correlated. It is worthwhile to note that only a Lorentzian $\chi_k$ leads to an efficient driving and that the resulting mode amplitudes decrease with frequency in contrast with the driving by the iron convection zone. Hence, it constitutes a seismic diagnostic to identify the driving by the convective core and also to determine the way the eddies are temporally-correlated, an essential constraint on  the dynamical properties of turbulent convection in cores of massive stars.  

\subsection{Mode amplitudes}
\label{ampl_core}

\begin{figure}[!]
\begin{center}
\includegraphics[height=6cm,width=9cm]{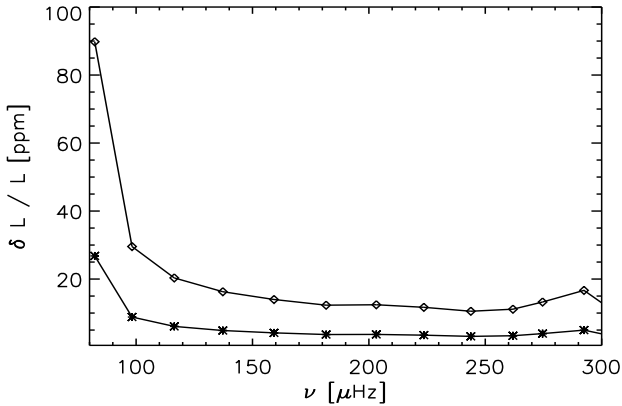}
\caption{Mode amplitude versus frequency, computed using the benchmark model (see Sect.~\ref{model_1D}) and a Lorentzian function to describe the eddy-time correlation function (see \eq{eqn:GF1}). The upper line (with diamond dots) corresponds to the upper limit for the injection length-scale that is. The lower line (with stars) corresponds to an injection length-scale that corresponds to the minimum of the pressure scale height in the convective core.}
\label{fig8}
\end{center}
\end{figure}

Mode amplitudes are presented in figure~\ref{fig8}. To investigate the sensitivity to the injection length-scale, we compute amplitude with an upper limit that is the convective core size and a lower limit, the minimum of the pressure length-scale in the convective core. 
In the more optimistic estimate, relative magnitudes are found to reach up to $90$ ppm for the lowest frequency mode and tens of ppm for higher frequency modes. 
Such a result demonstrates that excitation by the convective core can lead to amplitudes well above the CoRoT detection threshold ($\approx 1$ ppm) as well as the solar maximum ($\approx 3$ ppm). 
In the pessimistic case, the amplitude is near the CoRoT threshold, which makes the detection more difficult. 
Note also, that those results assume a Lorentzian description of the eddy-time correlation function ($\chi_k$), and that using another prescription such as an exponential decrease or a Gaussian one lead to very small amplitudes well below one ppm. 

Nevertheless, it is worthwhile to note that the frequency behaviour associated with the excitation by the convective core is very different from that obtained for the excitation by the iron convection zone. Hence, it constitutes a seismic signature that would be a powerful tool to identify the driving region.

\section{Concluding remarks}
\label{conclusion}

We performed an exploratory study of the driving (and damping) of acoustic modes by turbulent convection in a $10 \,  M_\odot$ star. 
It turns out that both the convective region associated with the iron opacity bump and the convective core are able to efficiently drive acoustic modes so that mode amplitudes are found above the CoRoT detection threshold ($\approx 1$ ppm) as well as the solar maximum ($\approx 3$ ppm). 

However, uncertainties associated with the computation of mode amplitudes are important and are related to the poor knowledge of the dynamical properties of those convective regions. For the iron convective region, the main uncertainties come from the convective velocities as well as the injection-lentgh scale. For the convective core, the way eddies are temporally-correlated is crucial since excitation occurs in the off resonance regime. 

Computation of mode amplitudes also permitted us to emphasise potential seismic diagnostics. First, driving by the iron convective region results in an oscillation of the energy injection rates (and mode amplitudes) versus mode frequencies. It has been shown that this behaviour is potentially useful to determine the extent and the depth of this convective region, as well as the dynamical properties of the driving (e.g., turbulent velocities, injection length-scales). Second, 
the way the eddies are temporally-correlated will determine the efficiency of the driving by the convective core and the frequency behaviour. Hence, it constitutes an opportunity to strongly constrain $\chi_k$.

The detection of solar-like oscillations in a massive star such as V1449 Aql \citep{Belkacem09b} and the determination of the mode parameter in this star are a promising way to identify the driving region, since both have very different signatures, and would give physical constraints to the still poorly known treatment of convective regions.
 The presence of high frequency power in SPB stars \citep{Degroote09b} would also help to constrain the nature of the driving zone since these stars have luminosities definitively below the threshold value for the presence of a convective zone associated with the iron peak \citep{Cantiello09}. It should however be kept in mind that microscopic diffusion coupled with radiation acceleration could induce an accumulation of iron in the iron peak. A convective iron zone could thus also be present in stars whose luminosity are slightly below the threshold.

\begin{acknowledgements} 
We are indebted to J. Leibacher for his careful reading of
the manuscript and to R. Samadi and M.J. Goupil for their 
helpful remarks. K.B. also 
acknowledges financial support 
through a postdoctoral fellowship from the 
“Subside f\'ed\'eral pour la recherche 2009”, University of Li\`ege. 

\end{acknowledgements}


\appendix
\end{document}